\documentclass{elsart}

\usepackage{amssymb}
\usepackage{amsmath}
\journal{Physics Letters A}

\begin{document}

\begin{frontmatter}



\title{Noiseless subsystems and Bell inequalities in curved spacetime}


\author{Jonathan L. Ball}

\address{Centre for Quantum Computation, Clarendon Laboratory, Department of Physics, University of Oxford OX1 3PU, U.K.}
\ead{j.ball1@physics.ox.ac.uk}

\begin{abstract}
We examine the use of noiseless subsystems for quantum information
processing between two parties who do not share a common reference
frame. In particular we focus on Bell inequalities in curved
spaces and outline a theoretical procedure to test a Bell
inequality, demonstrating the wide applicability of noiseless
subsystems.
\end{abstract}

\end{frontmatter}

The founding principle of quantum information theory is that
information is physical \cite{Landauer}. Far from being an
abstract quantity, information only exists in real physical
systems. As such, the behaviour of information itself is
intimately entwined with the nature of physical law governing the
universe.

The fruitful union of classical information theory and quantum
mechanics paved the way for the development of quantum
computation, quantum algorithms, quantum cryptography, quantum
teleportation and our understanding of entanglement \cite{ekert}.
However, the field of quantum information has for the most part
avoided many of the conceptual challenges borne from the insight
that spacetime is dynamical and curved. Some recent effort has
attempted to translate many of the aforementioned concepts to the
special relativistic setting (e.g. \cite{peresterno,terno2} ), and
an examination of the entanglement shared between modes of a
scalar field when one of the observers is uniformly accelerated
has been undertaken \cite{ivette}. Nevertheless, much further
analysis in the general relativistic setting is still required.

A natural mediator of quantum information is the photon, and much
attention is currently being directed towards the generation,
manipulation and measurement of photon states. Photons have been
employed in practical realizations of quantum cryptographic
protocols, quantum teleportation, fundamental tests of quantum
mechanics and may well play a major role in the structure of
quantum computers (see e.g. \cite{ekert}). The radiation field has
also played an important role in Relativity theory: indeed the
deflection of light from distant stars due to the curvature of
space-time around our sun provided an invaluable experimental
confirmation of the predictions of the General Theory (see e.g.
\cite{bookrel}).

Probing the structure of quantum mechanics is also crucially
important, a task made more difficult due to a number of
cointerintuitive features. Of great significance is the work of
Bell \cite{bellbook}. Bell's theorem is a general proof,
independent of any physical theory, arguing that there is an upper
limit to the correlation between distant physical events, provided
that one is prepared to accept the principle of local causes. This
theorem often takes the form of an inequality, the violation of
which implies that one of the assumptions inherent in its
derivation must be false.

A typical setup for considering Bell inequalities is the following
(see \cite{isham}). Suppose that two experimenters, Alice (A) and
Bob (B), reside at spacelike separated locations. A particle pair
source is located at O somewhere between them, and it periodically
creates a pair of particles, one of which travels to Alice and the
other to Bob. Each particle has some degree of freedom that may be
measured. Alice and Bob may make one of two possible measurements,
$A$ ($B$) or $A'$ ($B'$), and the result of either measurement is
$\pm 1$.  We shall assume hypothetically that the system under
consideration obeys the principles of Local Determinism and
Objective Reality. The first principle encompasses the assumption
that the results of Alice's measurements are locally determined by
the state of the particle she receives and not by the state of
Bob's particle. Furthermore, the result of her measurement ($\pm
1$) should be independent of the choice of Bob's measurement ($B$
or $B'$). The second assumption asserts that there is sufficient
information encoded in the internal state of each particle to
determine the outcome of either measurement made on it.  The
allowed possible outcomes of the measurements made by Alice and
Bob on the $n$th particle pair are therefore $A_n=\pm 1, A'_n=\pm
1, B_n=\pm 1, B'_n=\pm 1$. Local determinism and objective reality
allow us to assign a joint probability distribution $P(A,A',B,B')$
to the measurement outcomes. It is a simple matter to show that
the correlation function defined as
$g_n=A_nB_n+A'_nB_n+A_nB'_n-A'_nB'_n$ has value either $+2$ or
$-2$, and averaging over a number $N$ of measurement results it
follows that
\begin{equation}\frac{1}{N}|\sum^N_{n=1}g_n|\leq
2. \label{Eq:bellviolation}
\end{equation}
Suppose that we choose to test a Bell inequality using an
entangled quantum state of two spin-$\frac{1}{2}$ particles,
produced at O, of the form
\begin{equation}
\frac{1}{\sqrt{2}}(|0\rangle_{\mathcal{A}}|1\rangle_\mathcal{B}-|1\rangle_\mathcal{A}|0\rangle_\mathcal{B}),
\end{equation}
where $|1\rangle$ and $|0\rangle$ denote spin projections along,
for example, the $z$-axis, for example ($\mid\uparrow\rangle$ and
$\mid\downarrow\rangle$ respectively). Particle $\mathcal{A}$
propagates to Alice and particle $\mathcal{B}$ propagates to Bob.
Each party then selects axes along which to perform spin
projection measurements on their particle. Alice's measurements
correspond to $A=\hat{{\bf \sigma}}_z=|0\rangle\langle
0|-|1\rangle\langle 1|$ or $A'=\cos\phi\hat{{\bf
\sigma}}_z+\sin\phi\hat{{\bf \sigma}}_x$ where $\hat{{\bf
\sigma}}_x=|0\rangle\langle 1|+|1\rangle\langle 0 |$. Bob's
measurements, meanwhile, are either $B=\hat{{\bf \sigma}}_z$ or
$B'=\cos\phi\hat{{\bf \sigma}}_z-\sin\phi\hat{{\bf \sigma}}_x$.
Both observables have eigenvalues $\pm 1$.  A standard calculation
shows that $\frac{1}{N}|\sum_ng_n|=|1+2\cos\phi-\cos 2\phi |$, so
that the Bell inequality is violated for $0\leq\phi\leq \pi/2$.
Thus far, experimental tests of Bell inequalities strongly support
the predictions of Quantum Mechanics (see e.g. \cite{Aspect}).

However, the sharing of a reference frame to which Alice and Bob
may relate their spin projection measurements cannot be taken as a
preexisting component in a general communication scenario: instead
it must  be viewed as an expensive resource \cite{BartRudoPRL03}.
Indeed, if such an element is not given, then establishing perfect
alignment between two reference frames requires an infinite amount
of classical communication. The previous discussion pertains to
the case where the particles produced at the source propagate
through  flat spacetime since in this case the notion of a
particular direction in space, a necessary requirement for spin
projections along certain axes,  is the same for Alice and Bob.
For example, a statement such as \lq two particles located at
different space regions have opposite spin projections along the
$z$-axis ' has a definite meaning when the gravitational field is
negligible.  However, in reality it may well be the case that the
region of spacetime between two spacelike separated observers is
not entirely flat.

In general relativity, a gravitational field is represented by a
curved spacetime structure with metric $g_{\mu\nu}(x)$, which
implies a breakdown of the global rotational symmetry. However, a
spin represents the rotational symmetry of a system, and so can
only be satisfactorily defined locally by considering a local
inertial frame and invoking local rotational symmetry.  This is
achieved by making use of a vierbien $e^\mu_a(x)$ (inverse
$e^a_\mu(x)$) such that
\begin{equation}
e^\mu_a(x)e^\nu_b(x)g_{\mu\nu}(x)=\eta_{ab},
\end{equation}
where $\eta_{ab}$ is the Minkowski metric (metric signature
diag$(-1,1,1,1)$), and
\begin{equation}
e^a_\mu(x)e^\nu_a(x)=\delta^\mu_\nu\quad{ };\quad{
}e^a_\mu(x)e^\mu_b(x)=\delta^a_b.
\end{equation}
The vierbein represents a coordinate transformation from a general
coordinate system to a local one at each point in spacetime.  The
exact choice of local inertial frame is not unique, since an
inertial frame will remain inertial under Lorentz transformations,
and so the vierbein has the same degree of freedom, which is known
as the local Lorentz transformation. Using this local Lorentz
transformation, a particle with a given spin can be defined in
curved spacetime. The absence of a global timelike Killing vector
field in general curved spacetime makes the definition of a
particle somewhat difficult \cite{jif}. However in this case a
particle may be specified by making use of the vierbein
$e^\mu_0(x)$, which relates the local time coordinate to the
global time coordinate. A particle of spin-$\frac{1}{2}$ in a
region of curved spacetime is defined as a particle whose
one-particle states form a spin-$\frac{1}{2}$ representation of
the local Lorentz transformation (not the general coordinate
transformation). A particle of mass $m$ and four-momentum
$p^\mu(x)=m u^\mu(x)$ in the general coordinate system has
four-momentum defined by $p^a(x)=e^a_\mu p^\mu(x)$ in the local
coordinate system. The one-particle state in the local inertial
frame at $x^\mu$ is then specified by using the $z$ component of
spin $S_z \in\{\mid\uparrow\rangle, \mid\downarrow\rangle\}$, so
that the entire quantum state is $|p^a(x),S_z,x\rangle$. This
state does not represent a localized state at $x^\mu$ with
definite momentum $p^a(x)$, but instead an extended state with
momentum $p^a(x)$ when it is viewed in the local inertial frame
defined at spacetime location $x^\mu$.

A particle moving in a curved spacetime undergoes spin precession
due to two factors. The first is due to acceleration of the
particle by external forces, whilst the second is due to the
difference between local inertial frames defined at different
spacetime locations. The moving particle therefore undergoes a
succession of local Lorentz transformations.  These
transformations act on the spin by way of a succession of Wigner
rotations \cite{wigner}, the result of which is spin precession.

In \cite{ueda} both forms of spin procession are applied to a spin
singlet entangled state in the vicinity of a Schwarzschild black
hole. It is shown that acceleration and gravity serve to degrade
the anticorrelation in the spin directions that would be present
were the spacetime flat, and therefore the degree of violation of
a Bell inequality decreases. To obtain a maximal violation, Alice
and Bob must instead perform their spin measurements in
appropriately chosen \emph{different} directions. The exact
directions depend on the acceleration of the particles, the
spacetime metric and the locations of the observers. In
\cite{mensky1} the correlation between spin projections of two
particles created by the decay of a single scalar particle was
considered for the case where the particles propagate in a
gravitational field. Correlated directions are connected with each
other by performing parallel transport along the world lines of
the particles. Although this operation of parallel transport can
define equivalent directions in curved spacetime, the actual
directions for extracting maximal violation of a Bell inequality
are not in general the same as each other. Furthermore, in a
Schwarzschild geometry, the appropriate directions for maximal
violation depend so sensitively on the positions of the observers
that achieving such violation is rather challenging. In fact, a
tiny uncertainty in the position of the observers results in a
fatal error in quantum communication near the event horizon.

As such, testing Bell inequalities in curved spacetime using
spin-$\frac{1}{2}$ particles is subject to a number of significant
problems.  Instead, information may be encoded in the polarization
degree of freedom of a single photon.  In this case the states
$|0\rangle$ and $|1\rangle$ are mapped onto two orthogonal
polarization states, for example $\mid\leftrightarrow\rangle$
(horizontal polarization) and $\mid\updownarrow\rangle$ (vertical
polarization) respectively (a Bell inequality can then be tested
using a singlet state of two photons of the form
$\frac{1}{\sqrt{2}}(\mid\leftrightarrow\rangle\mid\updownarrow\rangle-\mid\updownarrow\rangle\mid\leftrightarrow\rangle)$).
Then in a general communication scheme, a message transmitted from
sender to receiver consists of a sequence of polarized light
pulses, each occupying a certain temporal slot. Information is
decoded by the receiver by measuring the polarization of the
light. Such schemes are subject to two dominant decoherence
mechanisms that can inhibit their performance. Firstly,  in a
multitude of physical systems the polarization of light undergoes
a random transformation during its journey from sender to
receiver, the result being that the input polarization state is
not deducible from that measured at the output. An example is that
of  polarization rotation due to fluctuating birefringence in
optical fibers as a consequence of environmental conditions such
as temperature fluctuations and mechanical strain (aeolian
vibrations). This decoherence appears to be a major obstacle when
implementing quantum information processing tasks using the
polarization degree of freedom. However, certain quantum states
can be used which are immune to the destructive effects of
specific types of decoherence. This is the case where the
environment couples identically to each quantum state without
distinguishing between them. Such a high degree of symmetry allows
one to identify whole sectors of the system's Hilbert space that
are unaffected by such decoherence.  The second significant
decoherence process is the linear loss of light amplitude during
the propagation process, for example due to absorption or
scattering. These linear losses are typically
polarization-independent and their effect can usually be negated
by carefully postselecting transmissions unaffected by photon
loss. It suffices therefore to concentrate on the first
decoherence mechanism.

A single use of a bosonic communication channel is the
transmission of a pair of bosonic modes with associated
annihilation operators $\hat{a}_{\leftrightarrow}$ and
$\hat{a}_{\updownarrow}$ pertaining to orthogonal polarizations.
Birefringence effects may be modelled as a random U(2)
transformation between the operators. An element of the U(2) group
is denoted by ${\bf \Omega}$: this element induces a unitary
transformation $\hat{U}({\bf \Omega})$ in the two-mode Fock space.
Consider now a quantum state of radiation distributed over $N$
temporal slots. Propagation of the entire state of radiation
$\hat{\rho}$ through spacetime where each slot is subject to the
same random depolarization results in the following transformation
\cite{hybridballban}
\begin{equation}
\hat{\rho} \mapsto \int_{\text{U(2)}} d{\bf \Omega} [\hat{U}({\bf
\Omega})]^{\otimes N}\hat{\rho} [\hat{U}^\dagger({\bf
\Omega})]^{\otimes N} \label{Eq:depol},
\end{equation}
where $d{\bf \Omega}$ denotes the invariant Haar measure in U(2).
$\hat{U}({\bf\Omega})$ may be decomposed as a product of the
overall phase factor $e^{-i\alpha({\bf \Omega})}$ and the
remaining SU(2) matrix ${\bf \Omega'}=e^{i\alpha({\bf
\Omega})}{\bf \Omega}$. However, when the communicating parties do
not share a common phase reference, the phase factor
$e^{-i\alpha}$ varies between consecutive uses of the channel, and
so only the SU(2) transformation ${\bf\Omega '}$ is constant. The
unitary transformation $\hat{U}({\bf\Omega})$ acting on a slot
containing $l$ photons may then be written in terms of
$\hat{\mathcal{D} }^{l/2}({\bf\Omega '})$ matrices which are
elements of the $(l+1)$-dimensional irreducible representation of
SU(2). These matrices satisfy the following group element
orthogonality relation:
\begin{equation}
\int\text{d}{\bf\Omega'} [ \mathcal{D}
^{j}_{mn}({\bf\Omega'})]^{\ast} \mathcal{D}
^{j'}_{m'n'}({\bf\Omega'}) = \frac{1}{2j+1} \delta_{jj'}
\delta_{mm'} \delta_{nn'}. \label{Eq:orthogprop}
\end{equation}
Performing the invariant integration in Eq.~(\ref{Eq:depol})
reveals sectors of the total system's Hilbert space which are
immune to the effects of collective depolarization and can
therefore be used to encode information in a robust manner.

The testing of Bell's theorem using certain sectors of the total
Hilbert space unaffected by noise is discussed in
\cite{cabello}(without resorting to inequalities). The resulting
protocol requires no reference frame alignments. In total, 8
photons are required, of which four are sent to Alice and four to
Bob. Generation of the states required is however rather
experimentally challenging.  We shall now present a rather
different protocol to test Bell inequalities using only 6 photons
(3 are sent to Alice and 3 to Bob). A practical implementation of
a scheme using only 6 photons would certainly be less sensitive to
photon losses than the previous case, either in optical fibers or
free space.

We shall now address the theoretical requirements for such a
protocol in more detail. Three photons may be prepared in a state
labelled by spin $j=3/2$ or spin $j=1/2$. We make use of the
following decomposition:
\begin{align}
\hat{\mathcal{D}}^{1/2}({\bf
\Omega'})\otimes\hat{\mathcal{D}}^{1/2}({\bf
\Omega'}&)\otimes\hat{\mathcal{D}}^{1/2}({\bf \Omega'})\nonumber\\
&=\hat{\mathcal{D}}^{3/2}({\bf \Omega'})
\oplus\hat{\mathcal{D}}^{1/2}({\bf
\Omega'})\oplus\hat{\mathcal{D}}^{1/2}({\bf \Omega'}).
\end{align}
The total density matrix describing the radiation field consisting
of 3 photons may be written as a direct sum of one $j=3/2$ and two
$j=1/2$ subspaces. Under the depolarization transformation,
Eq.~(\ref{Eq:orthogprop}) ensures that information encoded in the
$j=3/2$ subspace is not mixed with that in the $j=1/2$ subspaces,
and vice versa.  We shall concern ourselves with the two $j=1/2$
subspaces of the total Hilbert space.

Our new protocol involves the use of the following orthonormal
spin-$1/2$ states of 3 photons, introduced in \cite{KempeprAFTQC}
in the context of fault tolerant universal quantum computation:
\begin{eqnarray}
|0'_L\rangle= & \frac{1}{\sqrt{2}}(|010\rangle-|100\rangle),\nonumber\\
|0''_L\rangle= & \frac{1}{\sqrt{2}}(|011\rangle-|101\rangle),\nonumber\\
|1'_L\rangle= & \frac{1}{\sqrt{6}}(-2|001\rangle+|010\rangle+|100\rangle),\nonumber\\
|1''_L\rangle= &
\frac{1}{\sqrt{6}}(2|110\rangle-|101\rangle-|011\rangle),
\end{eqnarray}
where $|0_L\rangle$ and $|1_L\rangle$ denote logical \lq 0' and
\lq 1', and the $'$ and $''$ denote states within these
two-dimensional subsystems.  The general density matrix structure
for $j=1/2$ in this basis is
\begin{equation}
\label{Eq:basisstates}
\begin{pmatrix}
|0'_L\rangle\langle 0'_L|&|0'_L\rangle\langle
0''_L|&|0'_L\rangle\langle
1'_L|&|0'_L\rangle\langle 1''_L|\\
|0''_L\rangle\langle 0'_L|&|0''_L\rangle\langle
0''_L|&|0''_L\rangle\langle
1'_L|&|0''_L\rangle\langle 1''_L|\\
|1'_L\rangle\langle 0'_L|&|1'_L\rangle\langle
0''_L|&|1'_L\rangle\langle
1'_L|&|1'_L\rangle\langle 1''_L|\\
|1''_L\rangle\langle 0'_L|&|1''_L\rangle\langle
0''_L|&|1''_L\rangle\langle
1'_L|&|1''_L\rangle\langle 1''_L|\\
\end{pmatrix}.
\end{equation}
Under collective depolarization, the $|0'_L\rangle$ and
$|0''_L\rangle$ may be mixed amongst themselves but not with the
states $|1'_L\rangle$ and $|1''_L\rangle$, and vice versa.

Now suppose that a particle source residing at O produces the
following 6 photon state:
\begin{equation}
|\Psi\rangle=\frac{1}{\sqrt{2}}(|0'_L\rangle^{1-3}\otimes|1'_L\rangle^{4-6}-|1'_L\rangle^{1-3}\otimes|0'_L\rangle^{4-6}).
\end{equation}
Particles 1-3 propagate to Alice and particles 4-6 propagate to
Bob, who are spacelike separated parties. The input density matrix
for the 6 photon state has the explicit form
\begin{align}
\hat{\rho}=&\frac{1}{2}(|0'_L\rangle^{1-3}\langle
0'_L|\otimes|1'_L\rangle^{4-6}\langle
1'_L|\nonumber\\
& -|0'_L\rangle^{1-3}\langle 1'_L|\otimes|1'_L\rangle^{1-3}\langle
0'_L|\nonumber\\
&-|1'_L\rangle^{1-3}\langle 0'_L|\otimes|0'_L\rangle^{4-6}\langle
1'_L|\nonumber\\
& +|1'_L\rangle^{1-3}\langle 1'_L|\otimes|0'_L\rangle^{4-6}\langle
0'_L|)
\end{align}
and reduced density matrices
\begin{equation}
\hat{\varrho}^{1-3}_{\text{in}}=\hat{\varsigma}^{4-6}_{\text{in}}=\frac{1}{2}(|0'_L\rangle\langle
0'_L|+|1'_L\rangle\langle 1'_L|).
\end{equation}
Collective depolarization of this state takes the form
\begin{eqnarray}
\hat{\rho}& \mapsto & \int d{\bf \Omega}_A\int d{\bf \Omega}_B\nonumber\\
& &[\hat{U}({\bf \Omega}_A)]^{\otimes 3}[\hat{U}({\bf \Omega
}_B)]^{\otimes 3} \hat{\rho} [\hat{U}^\dagger({\bf
\Omega}_B)]^{\otimes 3}[\hat{U}^\dagger({\bf \Omega}_A)]^{\otimes
3},\nonumber\\
\end{eqnarray}
where $\hat{U}({\bf \Omega}_A)$ and $\hat{U}({\bf \Omega}_B)$,
which act on each of photons 1-3 and 4-6 respectively, are not
necessarily the same. The resulting output reduced density matrix
for each three photon state has the form
\begin{equation}
\hat{\varrho}^{1-3}_{\text{out}}=\hat{\varsigma}^{4-6}_{\text{out}}=\frac{1}{2}\begin{pmatrix}\frac{1}{2}&0&0&0\\
0&\frac{1}{2}&0&0\\
0&0&\frac{1}{2}&0\\
0&0&0&\frac{1}{2}\\
\end{pmatrix}.
\end{equation}
To implement a Bell inequality test we first require measurements
analogous to the  $\hat{{\bf \sigma}}_z$ and $\hat{{\bf
\sigma}}_x$ discussed earlier. Although the required measurements
no longer operationally measure spin direction, their role in a
Bell inequality test is the same. They may be implemented using
measurement operators with self-adjoint matrix representations (in
the basis given in Eq.~(\ref{Eq:basisstates}))
\begin{equation}
\hat{{\bf \sigma}}_z=\begin{pmatrix}1&0&0&0\\
0&1&0&0\\
0&0&-1&0\\
0&0&0&-1\\
\end{pmatrix}\quad{ };\quad{ }
\hat{{\bf \sigma}}_x=\begin{pmatrix} 0&0&1&0\\
0&0&0&1\\
1&0&0&0\\
0&1&0&0
\end{pmatrix}.
\end{equation}
We also require the implementation of measurements which are
linear combinations of such operators. A calculation identical in
spirit to that pertaining to spin projection measurements
reproduces exactly the predicted violation of the Bell inequality
in Eq.~(\ref{Eq:bellviolation}), yet this time we are able to use
noiseless subsystems constructed out of 3 photons, immune to the
destructive effects of collective depolarization and without
resorting to a shared reference frame.

The geodesics followed by massive test particles with non-zero
spin in an external gravitational field depend upon the
orientation of the spin.  If such an effect were to hold for
photons, then the result would be a polarization-dependent
deflection of light passing in the vicinity of the sun, or
polarization-dependent time delay of pulsar signals. In Einstein's
theory of gravity, it can be shown that there is no birefringence
and photons follow null geodesics irrespective of their
polarization.  Therefore, experimental observation of
gravity-induced birefringence would indicate new physics beyond
Einstein's gravity \cite{mohanty} or the standard model. This may
well be the case for plausible candidate theories unifying quantum
field theory with general relativity. Indeed, for a certain class
of more general theories, a notable prediction is that in a
gravitational field a pair of orthogonal linear polarization
states of light propagate with different phase velocities, a
violation of the Einstein Equivalence Principle. The phenomenon of
gravity-induced birefringence can in principle be measured and
therefore be used to impose constraints on the structure of these
more general theories.

Every theory of gravity which embodies the Einstein Equivalence
Principle necessarily includes the metric postulates, namely that
spacetime is endowed with a symmetric metric $g_{\mu\nu}$, that
the trajectories of freely falling bodies are geodesics of that
metric, and that in local freely falling reference frames, the
non-gravitational laws of physics are those of special relativity.
In every metric theory of gravity all non-gravitational fields
couple in the same way to a single second rank symmetric tensor
field. This is known as \emph{universal coupling}. Nonmetric
theories of gravity violate at least one of the metric postulates,
predicting more novel couplings between gravitational and
non-gravitational fields.

For example, in \cite{preuss} attention is given to static
spherically symmetric metric-affine fields and the effects of
solar torsion on light propagating from the Sun. It is shown that
a phase shift accumulates between the orthogonal polarization
components singled out by the stellar field as light propagates,
given by \cite{preussthesis}
\begin{equation}
\Delta\Phi=\sqrt{\frac{2}{3}}\frac{2\pi k^2\tilde{m}}{\lambda
R^2}\frac{(\mu +2)(\mu-1)}{(\mu +1)}
\end{equation}
where $k^2$ is a constant with the dimensions of length which
serves to characterize the strength of the coupling between the
torsion and the electromagnetic field, $\mu$ is the cosine of the
angle between the line of sight and normal on the stellar surface,
$\lambda$ is the wavelength of the light, $R$ is the radius of the
star and $\tilde{m}$ is the torsion mass. Although such
polarization rotations may be used to impose constraints on the
birefringence of spacetime, its existence evidently further
hinders any attempt to implement simple information transfer
protocols, such as testing Bell inequalities,  using polarized
light over such regions of spacetime. However noiseless subsystems
may be used to effectively combat such decoherence, allowing
reliable information transfer protocols and alignment-free tests
of Bell inequalities over vast spacetime distances as described.

Practical implementation of our protocol is experimentally
challenging. We are currently working on these issues
optimistically due to the possibility of implementing at least
some of the measurement operators using techniques such as single
photon measurements and parity checks \cite{lutkenhaus}.  Certain
properties of the states employed may well also prove useful in
practical applications. For example, exchanging the first and
second qubits of the states
$\frac{1}{\sqrt{2}}(|010\rangle-|100\rangle)$ and
$\frac{1}{\sqrt{2}}(|011\rangle-|101\rangle)$ results in the
picking up of an  overall phase factor of $-1$, whereas exchange
of the first and second qubits of the states
$\frac{1}{\sqrt{6}}(-2|001\rangle+|010\rangle+|100\rangle)$ and
$\frac{1}{\sqrt{6}}(2|110\rangle-|101\rangle-|011\rangle)$ has the
simple effect of acting as the identity.




\section*{Acknowledgements}

I would like to thank Konrad Banaszek, Frederic Schuller, Ivette
Fuentes-Schuller, Salvador Venegas-Andraca, Nikola Paunkovi\'{c},
Robert Spekkens, Daniel Browne and Norbert L\"{u}tkenhaus for
useful discussions and advice.






\end{document}